\documentclass[preprint,a4paper]{aastex63}

\usepackage{amsmath,amssymb}
\usepackage{natbib}
\usepackage{graphicx}
\usepackage{upgreek}

\newcommand{\lleft}{\left}
\newcommand{\rright}{\right}

\received{June 20, 2019}
\accepted{August 27, 2019}
\submitjournal{Solar Physics}

\shorttitle{Solar Filament Recognition Based on Deep Learning}
\shortauthors{Zhu et al.}

\begin{document}

\title{Solar Filament Recognition Based on Deep Learning}

\correspondingauthor{GaiFei Zhu}
\email{zhugaofei@nao.cas.cn}

\author[0000-0002-0405-7018]{GaoFei Zhu}
\affiliation{National Astronomical Observatories, Chinese Academy of Sciences, Beijing 100101, China}
\affiliation{University of Chinese Academy of Sciences, Beijing 100049, China}
\affiliation{Key Laboratory of Solar Activity, Chinese Academy of Sciences, Beijing 100101, China}

\author{GangHua Lin}
\affiliation{National Astronomical Observatories, Chinese Academy of Sciences, Beijing 100101, China}
\affiliation{Key Laboratory of Solar Activity, Chinese Academy of Sciences, Beijing 100101, China}

\author[0000-0003-4344-4093]{DongGuang Wang}
\affiliation{National Astronomical Observatories, Chinese Academy of Sciences, Beijing 100101, China}
\affiliation{Key Laboratory of Solar Activity, Chinese Academy of Sciences, Beijing 100101, China}

\author[0000-0002-1396-7603]{Suo Liu}
\affiliation{National Astronomical Observatories, Chinese Academy of Sciences, Beijing 100101, China}
\affiliation{Key Laboratory of Solar Activity, Chinese Academy of Sciences, Beijing 100101, China}
\affiliation{School of Astronomy and Space Sciences, University of Chinese Academy of Sciences, Beijing 100049, China}

\author[0000-0003-1675-1995]{Xiao Yang}
\affiliation{National Astronomical Observatories, Chinese Academy of Sciences, Beijing 100101, China}
\affiliation{Key Laboratory of Solar Activity, Chinese Academy of Sciences, Beijing 100101, China}

\begin{abstract}

The paper presents a reliable method using deep learning to recognize solar filaments in H$\upalpha$ full-disk solar images automatically. This method cannot only identify filaments accurately but also minimize the effects of noise points of the solar images. Firstly, a raw filament dataset is set up, consisting of tens of thousands of images required for deep learning. Secondly, an automated method for solar filament identification is developed using the U-Net deep convolutional network. To test the performance of the method, a dataset with 60 pairs of manually corrected H$\upalpha$ images is employed. These images are obtained from the Big Bear Solar Observatory/Full-Disk H-alpha Patrol Telescope (BBSO/FDHA) in 2013. Cross-validation indicates that the method can efficiently identify filaments in full-disk H$\upalpha$ images.

\end{abstract}


\section{Introduction} \label{sect:intro}

Solar filaments (also called prominences when they appear at the solar limb) are ones of the most obvious characteristics on the Sun. They are the projection of prominences on the solar surface, and look like elongated dark ribbons with irregular edges on the brighter solar disk. Solar filaments consist of relatively cool and dense gas suspended above the solar photosphere, generally lying along magnetic neutral lines \citep{LangKR2001}. Their temperatures and densities are much cooler and denser, respectively, than those of the surrounding corona. Furthermore, the eruptions of filaments are often associated with flares and coronal mass ejections (CMEs) \citep{Gilbert2000, Gopalswamy2003, JingJ2004, ChenPF2008, ChenPF2011, ZhangJ2012}. For example, with the improvement of observation, the eruptions of filaments show some typical characteristics of flares, and there are also flare ribbons and post flare loops after the eruptions of filaments \citep{Priest2002}. \citet{Forbes2000} points out that filament eruptions, flares, and CMEs can be regarded as the same physical process of release of magnetic energy in a different time and solar atmosphere height. Thus, it is crucial to study the evolution of solar filaments observationally and theoretically. The statistical analysis is equally important.

Since the rapid development of telescopes and computers, tremendous data has been created. Then the question of how to recognize filaments efficiently and automatically is raised, in particular on the data from H$\upalpha$ full-disk observations. Several automated detection algorithms have been proposed to recognize filaments in H$\upalpha$ images, concerning the finding of a threshold \citep{GaoJL2002, Shih2003, QuM2005, Fuller2005, YuanY2011, HaoQ2013}. \citet{Labrosse2010} have applied the support vector machine (SVM) to distinguish filaments from sunspots.

%f1 #&#
%
\begin{figure}[!htbp]
\centering
\includegraphics[width=0.85\textwidth]{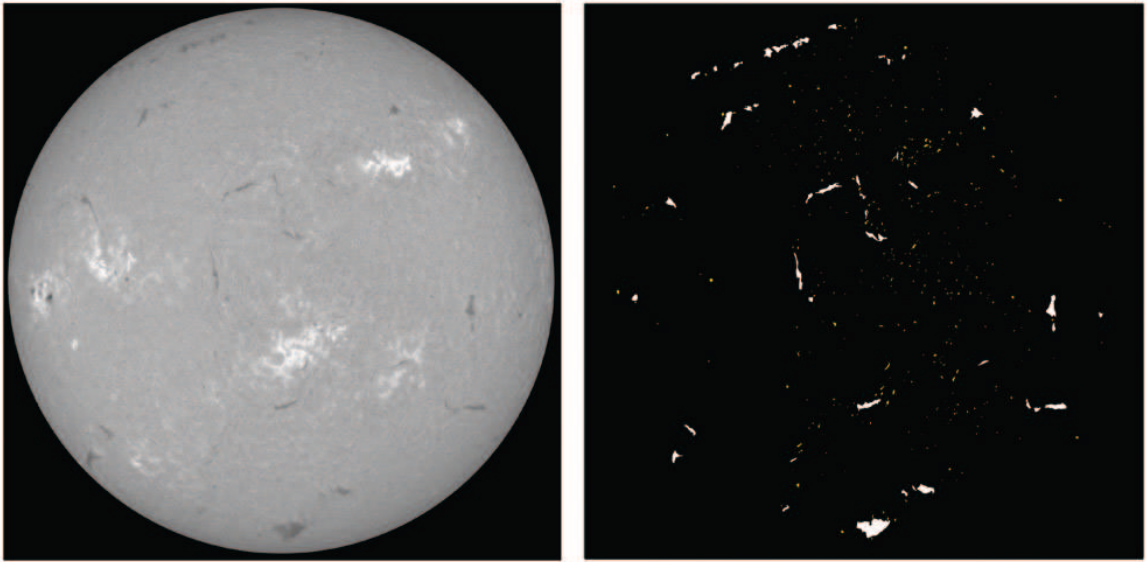}
\caption{One example of filament recognition using traditional image processing method. There are a lot of noise points in the solar disk, which are represented by the scattered and isolated yellow dispersion points in the right panel.}
\label{Fig1}
\end{figure}

We notice that the methods, as mentioned above, are mainly designed to find an optimal threshold, which will raise the following two problems. First, no matter what threshold is chosen, there will always be a large number of non-filament structures (noises or contaminations) (Figure~\ref{Fig1}) mixing in the recognition results, and another threshold should be set to remove them. This may rule out smaller filaments located around larger ones. In other words, there is no such a perfect threshold capable of removing noises and contaminations and keeping all filaments undamaged in H$\upalpha$ full-disk solar images. Second, when there are slightly uneven intensity distributions caused by the stable shutter
effect or other problems in the instrument, the above methods may mistakenly regard the dark contaminated areas as filaments. In that case, more methods need to be adopted to remove the uneven distribution on the solar disk.

\citet{Hinton2006} formally proposed the concept of deep learning with two viewpoints. Firstly, the features obtained from the deep learning networks can better reflect the essential attributes of the original data. Secondly, the results of upper training are used as initialization parameters in the lower training process, which can better solve the optimal problem. Once the method was implemented, it made a massive response in the academic circles. In particular, the team led by Geoffrey Hinton won the championship in the prestigious ImageNet Image Recognition Competition with the deep learning model -- AlexNet in 2012 \citep{Krizhevsky2012}. Since then, the algorithms for deep learning have achieved remarkable performance in many fields such as the medical, financial, art, and autopilot fields.

In this paper, we present a reliable algorithm using the improved U-Net, part of the deep convolutional neural networks, to recognize filaments for H$\upalpha$ full-disk solar images. The U-Net, first presented for biomedical image segmentation \citep{Ronneberger2015}, is a new approach to the study of filament segmentation in solar physics. We can identify the position of filaments and also accurately recognize them with few noise in the binary images. This paper is arranged as follows. The method of the improved U-Net network is described in Section~\ref{sect:method}. Moreover, in Section~\ref{sect:id}, we describe the details of the training process before filament identification. Then in Section~\ref{sect:eaa} are the experimental results. We draw our conclusions in the last section.

\section{Methods} \label{sect:method}

In this section, we introduce the fully convolutional networks (FCNs) proposed for semantic segmentation, then demonstrate the architecture of the improved U-Net network, in which we add several dropout layers in the network in order to enable it to work on the task of filament recognition. Moreover, the interpolation method is employed instead of deconvolution to reduce the training time.

%s2.1 #&#
\subsection{Fully Convolutional Networks}%
\label{sect:fcn}

Before the FCNs proposed by \citet{Long2015}, it was still a problem to realize pixel-level segmentation for dense end-to-end learning in an image. Most of the research focuses on image classification using the convolutional neural networks (CNNs, also referred to as ``convnet''). It is the first time that the full convolution layer is used to replace the fully connected layer to realize pixel-level image segmentation. In addition, Long's other significant contribution is to define the skip architectures that can combine the rough information from deep layers with the detailed information from shallow layers. Unlike CNNs, which classifies images, the FCNs can classify every pixel in an image to achieve image recognition.

\begin{figure}[!htbp]
\centering
\includegraphics[width=0.92\textwidth]{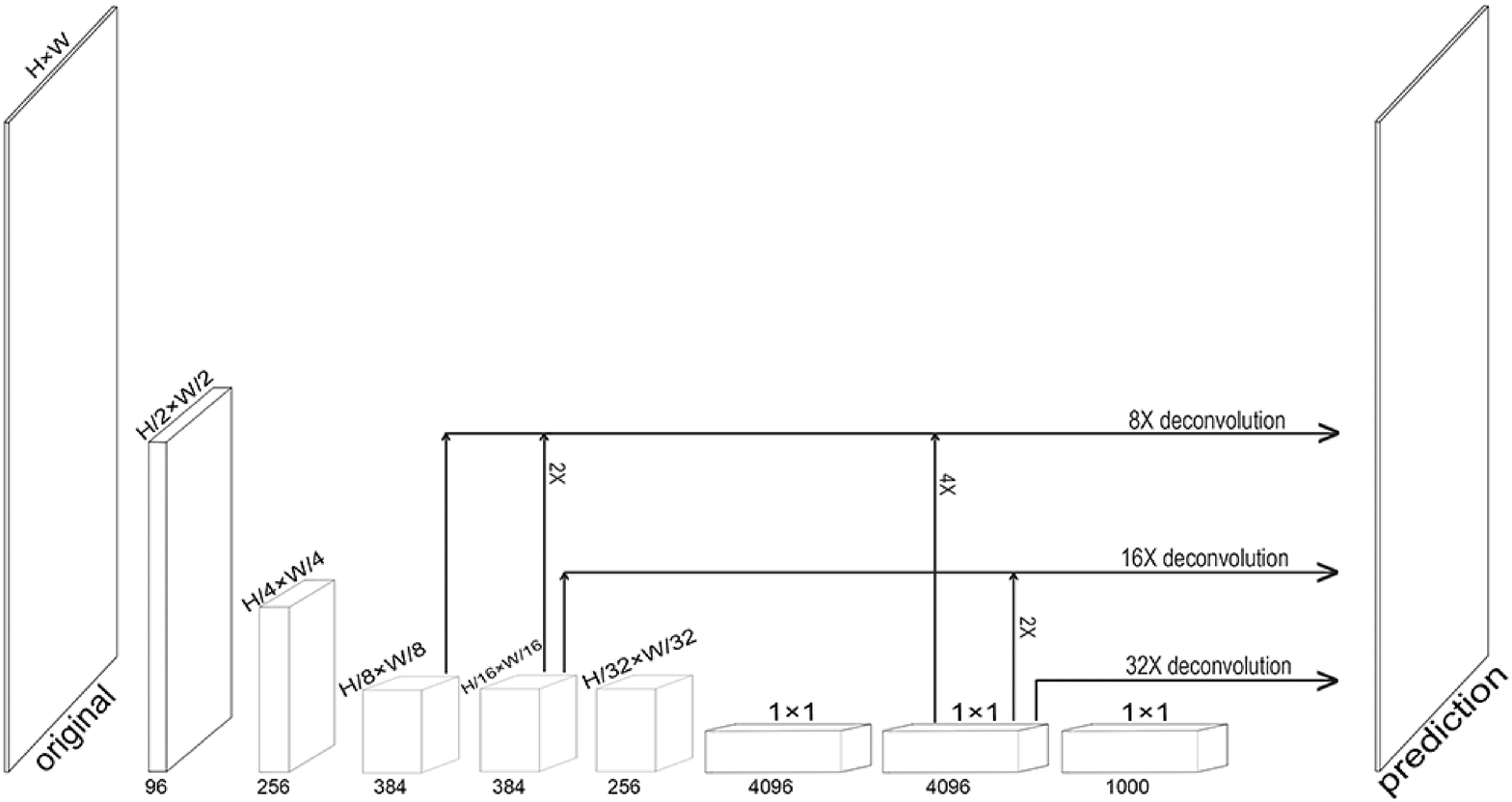}
\caption{Architecture of the fully convolutional networks (FCNs). The entire network is composed of convolutional layers without any fully connected layer. The FCNs combine coarse and high layer information with fine and low layer information.}
\label{Fig2}
\end{figure}

The FCNs modify the classical CNNs classification network \citep{Krizhevsky2012} consisting of five convolution layers and three full connected layers into fully convolutional networks (Figure~\ref{Fig2}). A basic convnet consists of convolution, pooling, and activation functions. The purpose of convolution is to extract feature maps from the prior layer by the learned convolution kernels. The activation function ``Rectified Linear Unit'' (ReLU) \citep{Nair2010} is used to add non-linear elements to the feature maps. The position element $s(i,j)$ in the feature map can be computed by
%
%e1 #&#
%
\begin{equation}
s(i,j) = \mathrm{ReLU} \Biggl\{ {\sum_{k = 1}^{N}
{({X_{k}} \otimes{W_{k}}) (i,j) + b} } \Biggr\} ,
\end{equation}
where $\mathrm{ReLU}(x) = \max(0,x)$, $N$ is the total channel number of input layer, $X_k$ is the $k$th input channel, $W_k$ is the $k$th convolution kernel of one of the input channels, and $b$ represents the adjustable bias. The basic function, ``ReLu'', is to make those negative values to 0. The convolution layer is aimed at extracting multi-dimensional feature maps from the previous layer.

%f2 #&#
%

In order to obtain more prominent feature information, the resolution of feature maps is compressed in the pooling layer under the effect of translation invariance. Generally, there are two kinds of pooling operations, ``average'' pooling and ``max'' pooling. The FCNs chooses the latter as its pooling strategy. The value at location $(i, j)$ in the $l$th pooling layer can be denoted
%
%e2 #&#
%
\begin{equation}
a_{ij}^{l} = \max\bigl(a_{mn}^{l - 1}
\bigr),\quad i \le m,n \le i + 2,
\end{equation}
where $m$ ($n$) is the overlapped region of pooling kernel in the $(l-1)$th convolution layer. The size of the pooling kernel is $3\times 3$ with a stride of 1. The pooling layer mainly reduces the size of the feature map of the upper layer and reduces the computational complexity of the network.

Although the FCNs can efficiently learn to make a dense prediction for per-pixel tasks like semantic segmentation, the original FCNs are not perfect enough because their segmentation results are still rough. Therefore, lots of studies have been done to improve the precision of image segmentation. Generally, there are two kinds of work dealing with this problem.

The first is to add dilated convolution behind the standard convolution layer to avoid losing information in the pooling process \citep{Yu2016}, and to improve the dilated convolution such as atrous spatial pyramid pooling \citep{ChenLC2014}, and fully connected conditional random fields (CRFs) \citep{ChenLC2016}.

The second is to build up the skip connections between the max-pooling layers and the up-sampling layers, for example, the DeconvNet \citep{Noh2015}, SegNet \citep{Badrinarayanan2017}, and the U-Net \citep{Ronneberger2015}. Due to the high efficiency and accuracy of the U-Net, we plan to apply it to do the task of segmentation of solar filaments. Also, it can be further improved to achieve better performance.

%s2.2 #&#
\subsection{Improved U-Net Based Filament Segmentation}%
\label{sect:iunet}

The U-Net is based on the FCNs. The developers modify and expand the network framework so that it can use very limited quantity of training images to obtain more accurate recognition results. Its architecture consists of a contracting path in which the dimensions of the feature maps are reduced due to the max-pooling, and an expansive path in which feature maps are combined with up-sampling maps at corresponding location \citep{Ronneberger2015}. There are several down-sampling and up-sampling blocks in the network.

%f3 #&#
%
\begin{figure}[!htbp]
\centering
\includegraphics[width=0.92\textwidth]{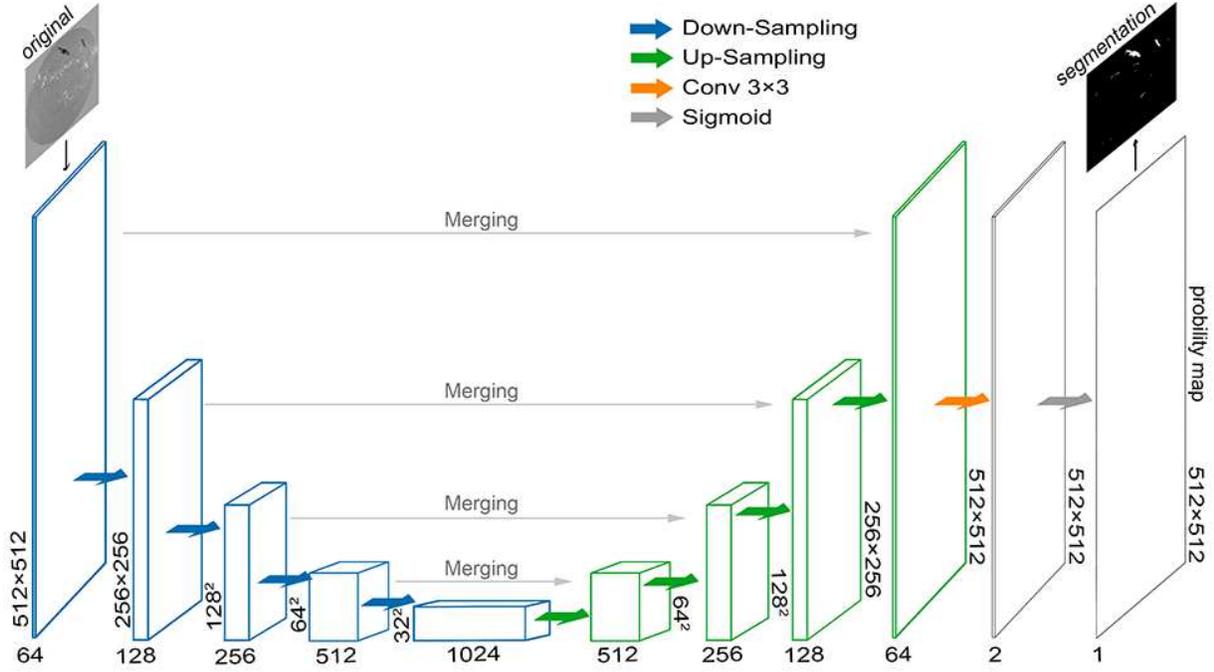}
\caption{Architecture of the improved U-Net network. It contains a contracting path shown with blue and an expansive path with green. The number below each box represents the number of channels. The size of feature maps is provided at the left and right sides of the box. Each box contains more operations as shown in Figure~\ref{Fig4}.}
\label{Fig3}
\end{figure}

Usually, filaments occupy quite small areas on the solar disk, and it is difficult to get accurate segmentation to the filaments using U-Net directly. Therefore, based on classic U-Net, we introduce the dropout layers \citep{Srivastava2014} behind the two convolutional layers in the first four down-sampling blocks. Additionally, we employ up-sampling operation with the function of the nearest-neighbor interpolation instead of deconvolution operation in the up-sampling blocks (Figure~\ref{Fig3}).

%f4 #&#
%
\begin{figure}[!htbp]
\centering
\includegraphics[width=0.7\textwidth]{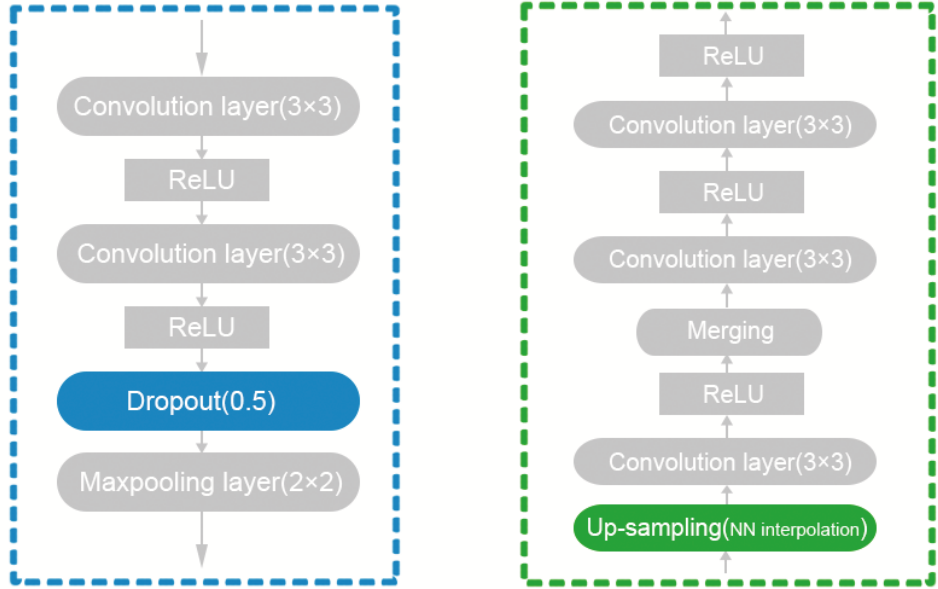}
\caption{Architecture of each colored box as shown in Figure~\ref{Fig3}. The blue dashed box contains two convolution layers, one dropout layer, and one max-pooling layer. The green dashed box consists of an up-sampling layer, a convolutional layer, a merging layer, and two convolutional layers. The size of convolution kernels is $3\times 3$. Each convolution layer is followed by an activation function. The dropout layer discards neurons at a 50\% probability. The up-sampling layer use the nearest-neighbor interpolation instead of deconvolution. Merging operation is used to connect the second convolution layer in each down-sampling box and the up-sampling layers in each corresponding up-sampling box.}
\label{Fig4}
\end{figure}

The first down-sampling block contains two convolution layers with the activation function of the ``ReLU'' and one max-pooling layer (Figure~\ref{Fig4}). In the first two convolution layers, there are 64
convolution kernels, the size of which is $3\times 3$ and the step size is 1. In the following down-sampling blocks, the number of convolution kernels is twice that of the former, and the size of kernels remains unchanged. To prevent over-fitting, we add several dropout layers after two convolution layers in the first four down-sampling blocks (Figure~\ref{Fig4}). The formula of the dropout layer is given by
%
%e3 #&#
%
\begin{equation}
\tilde{y}^{l} = B(p) \times{y^{l}},
\end{equation}
where $p$ is the probability of discarding neurons, generally being set as 0.5. $B$ function randomly generates 0 or 1. The neuron $y_{i}$ in the $(l+1)$th layer can be calculated as
%
%e4 #&#
%
\begin{equation}
y_{i}^{ ( {l + 1}  )} = f \bigl\{ {w_{i}^{(l + 1)}{{
\tilde{y}} ^{l}} + b_{i}^{(l + 1)}} \bigr\} ,\quad i
= 1,2, \ldots,N.
\end{equation}
It is designed to discard each neuron at a 50\% probability. 

In the up-sampling blocks, we use nearest-neighbor interpolation to resize image instead of deconvolution in order to guarantee the results to be reliable and at the same time, improve the speed of training (Figure~\ref{Fig4}). Because of the representativeness of features, the reliability comes from the approximate adjustment of the former feature layer by interpolation. The width and height of the original image are $w_{1}$ and $h_{1}$, and those of the scaled image are $w_{2}$ and $h_{2}$, respectively. The coordinates of the scaled image can be computed as:
%
%e5 #&#
%
\begin{equation}
\lleft\{ %
\begin{array} {l} w = \dfrac{{{w_{1}}}}{{{w_{2}}}},\qquad{w_{2}}
\ne0,
\\  \noalign{\vspace{4pt}}
h = \dfrac{{{h_{1}}}}{{{h_{2}}}},\qquad {h_{2}} \ne0,\qquad
\\   \noalign{\vspace{4pt}}
{x_{0}} = {\mathop{\mathrm{int}}} (x*w),
\\  \noalign{\vspace{4pt}}
{y_{0}} = {\mathop{\mathrm{int}}} (y*h), \end{array} %
\rright.
\end{equation}
where $(x_0,y_0)$ in the scaled image is equal to $(x,y)$ in the original image.

\section{Implementation Details} \label{sect:id}

%s3.1 #&#
\subsection{Data Preprocessing}%
\label{sect:dp} 

Since there is no common training sets available for filaments segmentation so far, we built up a filament segmentation dataset\footnote{The filament segmentation dataset is available in \url{http://sun.bao.ac.cn/hsos\_data/download/filaments-unet-dataset/}.} containing tens of thousands of images. The dataset includes two parts: the full-disk H$\upalpha$ solar images and the ground-truth maps. For filament recognition, the ground-truth maps refer to the correct segmentation results map for the supervised learning. They are binary maps, where 1 represents the filament region and 0 refers to the non-filament point. However, to obtain better segmentation results, further manual correction for the ground-truth maps is required.

%t3.1 #&#
%
\begin{table}[!htbp]
\centering
\caption{Parameters of data enhancement of the improved network.}
\label{mbh}%
\begin{tabular*}{0.6\textwidth}{ll}
\hline
\hline
Methods           & Range                                         \\
\hline
Rotation          & 0.2$^{\circ}$                                 \\
Shift             & 0.5 on both horizontal and vertical direction \\
Shear             & 0.05                                          \\
Zoom              & 0.05                                          \\
Flip horizontally & 50\% probability                              \\
Flip vertically   & 50\% probability                              \\
Fill mode         & Nearest                                       \\
\hline
\end{tabular*}
\end{table}

We selected 30 typical pairs of preprocessed and ground-truth images as original training set from tens of thousands of images, 20 pairs of them are chosen as the validation set and 10 pairs of them chosen to test the performance of the proposed method. Limb darkening is removed from the original images to avoid introducing too much noises and other undesirable factors. All the pixels outside the solar disk are set to gray. The raw images are labeled as ground truth using Photoshop software and traditional digital image processing.

Data augmentation is also critical for network invariance and robustness when a small number of training images are available (Table~\ref{mbh}). Flipping, shifting, and rotation invariance are the main methods. In order to obtain a usable model, only those images with distinct features are selected as training samples. Finally, 6040 training samples are generated from 30 high-quality H$\upalpha$ full-disk solar images.

The proposed method takes H$\upalpha$ full-disk solar images and ground-truth maps as inputs that are resized to $512\times 512$ and sent to the improved U-Net architecture to generate an appropriate weight model, which is used to segment filaments in H$\upalpha$ images. The improved U-Net network can train end-to-end even for a minimal number of images at high speed.

%s3.2 #&#
\subsection{Training}%
\label{sect:t}

Due to the hardware limitation of the experiment, the input tiles are scaled to $512\times 512$. For GeForce 1070Ti GPU, the memory will be exhausted if the batch size is set at 4 or greater. Therefore, it has to be set at 2. We choose ``Adam'' as the optimizer \citep{Kingma2014}, with the learning rate of 0.0001, beta1 of 0.9, and beta2 of 0.999. All weights are initialized by the ``He'' normal initializer \citep{HeK2015} with the mean value of 0 and the standard deviation of $\mathrm{{sqrt(2 / fan\_in)}}$, and all biases are initialized as 0.

In the task, the U-Net divides the solar full-disk into filament regions and non-filament regions. The sigmoid function $S(x)$ is applied to sort out the results, which is a kind of logistic function that converts all the results to probabilities within $(0,1)$. Its expression is
%
%e6 #&#
%
\begin{equation}
S(x) = \frac{1}{{1 + {\mathrm{e}^{ - x}}}},
\end{equation}
where $x$ is the processed result of the aforementioned network. The closer its value is to 1, the more likely the pixel is the target object.

The binary cross-entropy is the loss of the network,
%
%e7 #&#
%
\begin{equation}
\label{eq7}
{\mathrm{loss}} = - \sum_{i = 1}^{n} {{{\hat{y}}_{i}}\log
{y_{i}} + (1 - {{\hat{y}}_{i}})\log(1 - {y_{i}})} ,\quad i = 1,2, \ldots,N,
\end{equation}
where $y_{i}$ is the prediction result and ${\hat{y}_{i}}$ is the ground truth.

Without using a pre-trained model, our method can use the GPU to train a remarkable model in about 49.3 minutes. For the CPU, it takes about 57.1 hours to train an available model.

\section{Experiments} \label{sect:eaa}

%s4.1 #&#
\subsection{Performance Evaluation}%
\label{sect:pe}

The dice similarity coefficient (DSC), a measure of similarity between two binary sets, and the True Positive Rate (TPR) from common semantic segmentation evaluation strategy was measured in our filament segmentation. For the detected filaments, DSC provides the overlap measurement between marked regions in ground-truth map and the segmentation results, which is
%
%e8 #&#
%
\begin{equation}
\mathrm{DSC} = \frac{2\mathrm{TP}}{2\mathrm{TP} + \mathrm{FP} + \mathrm{FN}},
\end{equation}
where TP, FP and FN denote the true positive, false positive, and false negative measurements, respectively.

Additionally, the TPR measurement is as follows:
%
%e9 #&#
%
\begin{equation}
\mathrm{TPR} = \frac{\mathrm{TP}}{\mathrm{TP} + \mathrm{FN}},
\end{equation}
and the false positive rate (FPR) is
%
%e10 #&#
%
\begin{equation}
\mathrm{FPR} = \frac{\mathrm{FP}}{\mathrm{FP} + \mathrm{TN}}.
\end{equation}
The training loss and the binary accuracy are also showed in Figure~\ref{Fig7}, which describes the training process of this deep neural network. The binary accuracy is an evaluation metric to the accuracy of binary classification problems (Equation~\ref{eq7}).

%s4.2 #&#
\subsection{Experimental Results}%
\label{sect:er} 

\begin{figure}[!htbp]
\centering
\includegraphics[width=0.72\textwidth]{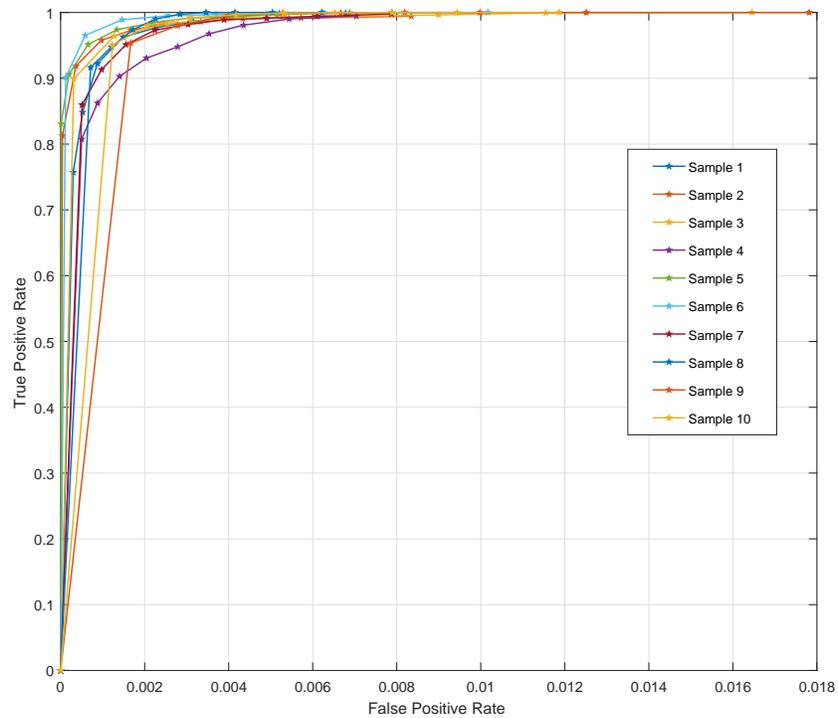}
\caption{ROC curve of the proposed network for ten testing samples (represented by different colors).}
\label{Fig5}
\end{figure}

We select ten pairs of images (including ten original images and ten ground-truth images) containing various shapes of filaments as our samples to evaluate the performance of the proposed network architecture. As the segmentation result is a probability map, different probability thresholds will have different effects on the final segmentation accuracy. The receiver operating characteristic (ROC) \citep{Fawcett2006} analysis, which is often used to evaluate the pros and cons of a binary classifier, is performed by plotting TPR and FPR at various threshold settings. So, we use the ROC curve to judge the quality of the segmentation and determine an appropriate probability threshold. We sort the non-zero probability values in each probability map incrementally and then divide them into ten groups. The maximum value of each group of probability values is used as the candidates' probability thresholds to obtain 10 groups of filament segmentation results for each probability map. In Figure~\ref{Fig5}, according to TPR and FPR calculated between segmentation results and ground truths, the ROC curves have been drawn for the selected test set. For each probability map, the point close to the upper left corner of the ROC plot is the point of the appropriate probability threshold. As shown in the figure, our method can achieve very low FPR and very high TPR. If the manual corrections of ground truth in the training set are more accurate, the accuracy of segmentation might be further improved. Theoretically, when the predicted probabilities of these small regions around large filaments are below the appropriate probability thresholds on the ROC curve, they would be neglected. However, it should be noted that these small regions with lower intensity (higher predicted probability) can still be recognized.
%
%f5 #&#
%

Table~\ref{tab2} lists not only the DSC results but also the TPR and the FPR that are generated from Figure~\ref{Fig5} when the proper probability thresholds are chosen. The probability thresholds that usually located in the upper left corner in Figure~\ref{Fig5} must satisfy the requirements of FPR as small as possible while those of TPR as large as possible. The position close to the upper left corner of Figure~\ref{Fig5} is the proper probability threshold, which makes the segmentation result the best. In the 10 samples, the highest TPR reaches 0.9642, while the corresponding lowest FPR is as low as 0.0002. The averaged TPR is 0.9145, and the averaged DSC is 0.8944. These show that our network is a viable strategy for solar filament recognition. We also find that the larger filaments are, the higher is the segmentation accuracy. On the contrary, the more scattered small filaments are, the lower is the segmentation accuracy. This is because, for large filaments, those apparent features can be detected easily.

\begin{table}
\centering
\caption{Performance indicators of TPR, FPR and DSC. PThr represents the appropriate probability threshold.}%
\label{tab2}
\tabcolsep=+3.8pt
\begin{tabular*}{\textwidth}{lllllllllll}
\hline
\hline
     & Sample1 & Sample2 & Sample3 & Sample4 & Sample5 & Sample6 & Sample7 & Sample8 & Sample9 & Sample10 \\
\hline
TPR  & 0.8482  & 0.9186  & 0.9642  & 0.8622  & 0.9060  & 0.9653  & 0.8597  & 0.9165  & 0.9535  & 0.9504   \\
FPR  & 0.0005  & 0.0003  & 0.0013  & 0.0008  & 0.0002  & 0.0006  & 0.0005  & 0.0007  & 0.0017  & 0.0012   \\
DSC  & 0.8818  & 0.9152  & 0.8904  & 0.8799  & 0.9400  & 0.9542  & 0.8876  & 0.8527  & 0.8282  & 0.9144   \\
PThr & 0.4838  & 0.3037  & 0.3679  & 0.5317  & 0.4769  & 0.5182  & 0.7238  & 0.6811  & 0.6470  & 0.7586   \\
\hline
\end{tabular*}
\end{table}

\begin{figure}[!htbp]
\centering
\includegraphics[width=0.5\textwidth]{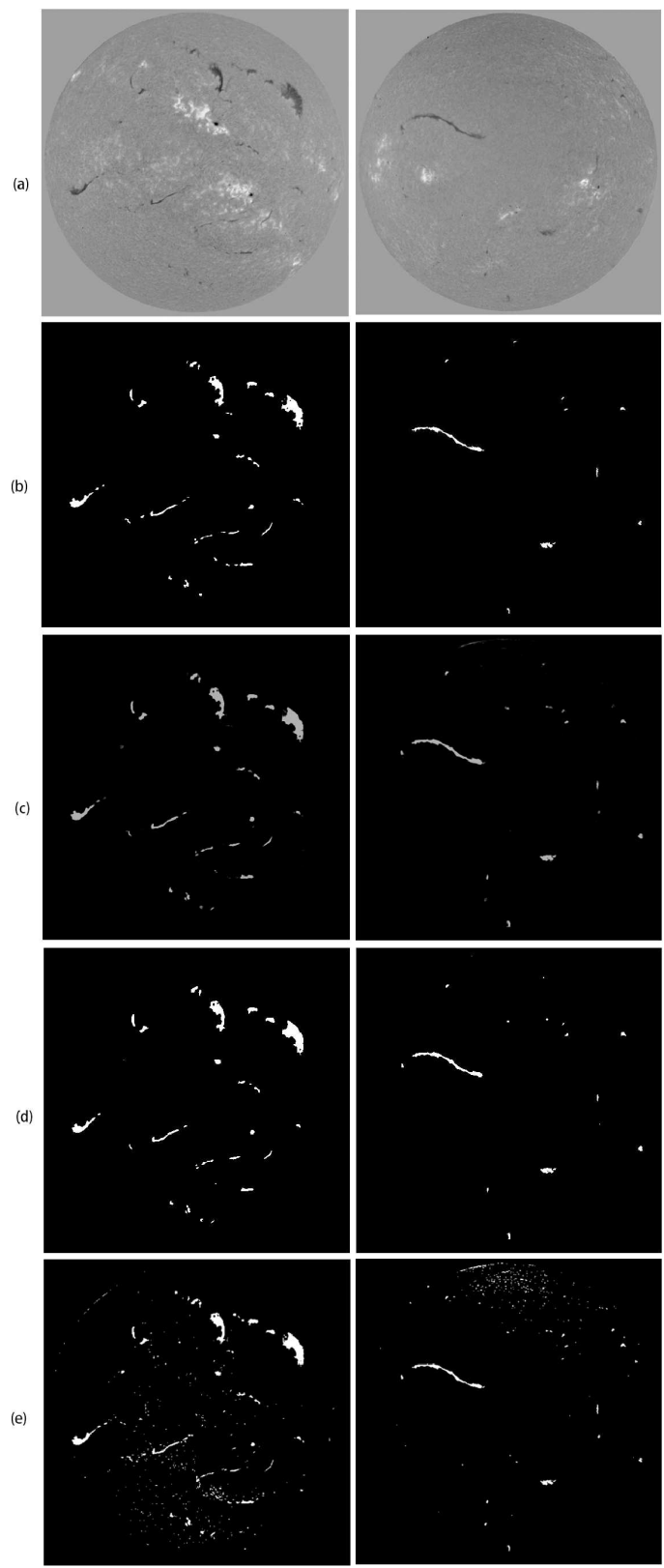}
\caption{One example of segmentation results using the improved U-Net network. (\textbf{a}) The preprocessed H$\upalpha$ images. (\textbf{b}) The ground truths. (\textbf{c}) The probability maps. (\textbf{d}) The segmentation results. (\textbf{e}) The segmentation result using traditional image processing.}
\label{Fig6}
\end{figure}

Figure~\ref{Fig6} shows some segmentation results compared with the ground truth. Meanwhile, the results generated with the traditional digital image processing are also compared with those with the proposed method. As is shown in Figure~\ref{Fig6}(e), the traditional image processing method produces numerous noise points that have to be removed in subsequent operations, which consequently may also cause the removal of some fragmented parts around the filaments at the same time. The results show that our method is a viable method for filament segmentation in full-disk H$\upalpha$ solar images.

%f6 #&#
%

\begin{figure}[!htbp]
\centering
\includegraphics[width=0.92\textwidth]{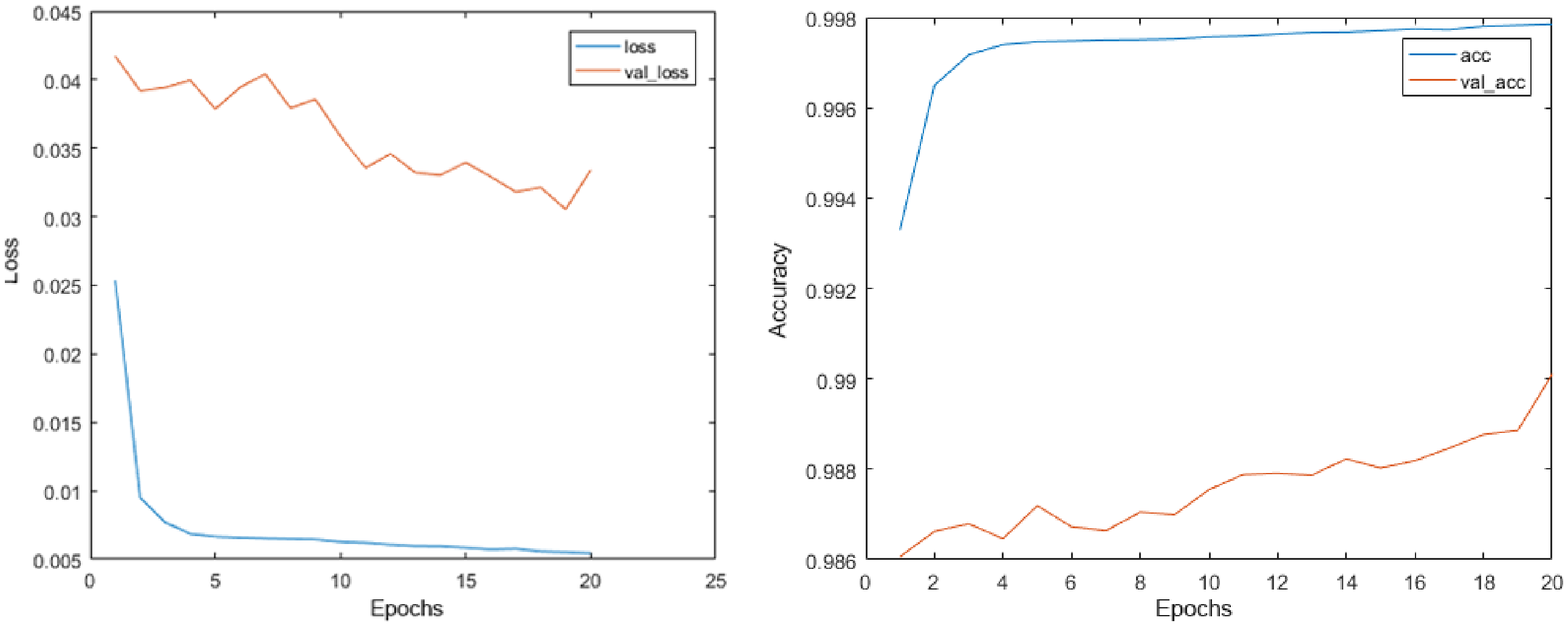}
\caption{Loss and accuracy in the training process of the proposed network using a training set.}
\label{Fig7}
\end{figure}

Figure~\ref{Fig7} shows the influence of training epochs on the segmentation performance of the proposed method that uses images of the training set and validation set. As is shown in the left panel, the loss of the training set of the proposed network architecture converges steeply at the first five epochs and settles at 0.005 after 20 epochs. The loss of the validation set shows an overall downward trend. For the right panel, the accuracy of the training set increases rapidly and is up to 0.997 at the epoch of 5. It settles at about 0.998 after 20 epochs. The accuracy of validation set shows an overall upward trend. Therefore, it is reasonable to set training epochs at 20 in the training process of the proposed network.

%f7 #&#
%

\section{Conclusions} \label{sect:c}

In this paper, we present a fully automated filament detection and segmentation method for H$\upalpha$ full-disk solar images using the improved U-Net deep convolution networks.\footnote{The code and model are available on GitHub (\url{https://github.com/GF-Zhu/Filament-Unet}).} A~dataset consisting of tens of thousands of images is available online. We demonstrate that our method can provide efficient segmentation compared with the semi-manual processed ground truth. In addition, our improved U-Net network can segment filaments directly and avoid generating segmentation with a large number of noise points. The proposed network can take less than an hour to get a usable model.

Even though we have achieved an inspiring result, there are still some limitations in the current work. First, for the images with obviously uneven intensity on the solar disk, the model may recognize those locations with too low intensity as filaments. The intensity features of these locations have strong similarities with typical filaments, which may lead to the misrecognition of the network. So, expanding network depth or increasing the diversity of the training set may be effective to solve the problem. Secondly, the network has the risk of falling into local optimum. In that case, we have to retrain the model. A fixed learning rate may be the cause of the problem. Adaptive adjustment to the learning rate is a possible strategy to address this kind of problem, which is our future research direction.

\acknowledgements

The work was funded by National Science Foundation of China (Grant Nos: U1531247 and 11427901), the 13th Five-year Informatization Plan of Chinese Academy of Sciences (Grant No. XXH13505-04), and the special foundation work of the Ministry of Science and Technology of China (Grant No: 2014FY120300). We thank BBSO for providing full-disk H$\upalpha$ images for the experiment.

%\bibliographystyle{aasjournal}
%\bibliography{zgf19}{}

\end{document}